\documentclass{pasj00}

\begin{document}
\SetRunningHead{T. Kato et al.}{Eclipsing Dwarf Nova GY Cancri}

\Received{}
\Accepted{}

\title{Outburst Photometry of the Eclipsing Dwarf Nova GY Cancri}

\author{Taichi \textsc{Kato}, Ryoko \textsc{Ishioka}, Makoto \textsc{Uemura}}
\affil{Department of Astronomy, Kyoto University,
       Sakyo-ku, Kyoto 606-8502}
\email{(tkato,ishioka,uemura)@kusastro.kyoto-u.ac.jp}



\maketitle

\begin{abstract}
   We observed the ROSAT-selected eclipsing dwarf nova GY Cnc
(=RX J0909.8+1849) during the 2001 November outburst.  We refined
the orbital period to be 0.17544251(5) d.  The fading portion of
the outburst was indistinguishable from those of typical dwarf novae
with similar orbital periods.  However, the signature of orbital humps
(or a hot spot) was far less prominently observed in the orbital light
curves and eclipse profiles than in usual dwarf novae with similar
orbital periods.  The combination of low frequency of outbursts and
the apparent lack of slowly rising, long outbursts in GY Cnc is difficult
to reconcile within the standard framework of dwarf novae.
We suspect that GY Cnc may be the first above-the-gap counterpart of
unusual eclipsing dwarf novae HT Cas and IR Com.
\end{abstract}

\section{Introduction}

   Cataclysmic variables (CVs) are close binary systems consisting of
a white dwarf and a red-dwarf secondary transferring matter via
Roche-lobe overflow.  Dwarf novae are a class of CVs showing outbursts,
which are believed to be a result of instabilities in the accretion disk
[see \citet{osa96review} for a review].

   Depending on orbital inclination, some dwarf novae show a various
degree of eclipses.  Eclipses in such dwarf novae provide a powerful tool
in studying the time-variation of the structure of accretion disks
(e.g. \cite{EclipseMapping}; \cite{woo89ippeg}).

   GY Cnc (=RX J0909.8+1849 = HS 0907+1902)
is a CV identified in the course of the Hamburg/RASS identifications of
ROSAT sources \citep{bad98RASSID}.  The dwarf nova nature was suspected
upon the recognition of an apparent outburst on Guide Star Catalog (GSC)
\citep{kat00gycnc}.  During the 2000 February outburst, several
observers independently discovered that the object shows deep eclipses
(\cite{kat00gycnc} and references therein; \cite{gan00gycnc}).
\citet{tho00gycnc} further studied the object spectroscopically,
and obtained component masses.  \citet{tho00gycnc} reported that
a modeling with a flat, Keplerian disk did not yield a good fit
to the observed profile of the quiescent H$\alpha$ emission line.
\citet{sha00gycnc} studied the object during post-outburst quiescence,
and derived orbital parameters.  \citet{sha00gycnc} found that the
bright spot (hot spot) is less conspicuous in GY Cnc compared to
IP Peg, the eclipsing dwarf nova having similar orbital parameters.
These observations suggest a some degree of peculiarity in GY Cnc
compared to other dwarf novae above the period gap.

   In spite of independent eclipse detections, the observation of the
2000 February outburst turned out to be rather fragmentary.  We present
first-ever complete photometric coverage of the declining branch of
the 2001 November outburst.

\section{Observation}

\begin{figure}
  \begin{center}
    \FigureFile(88mm,60mm){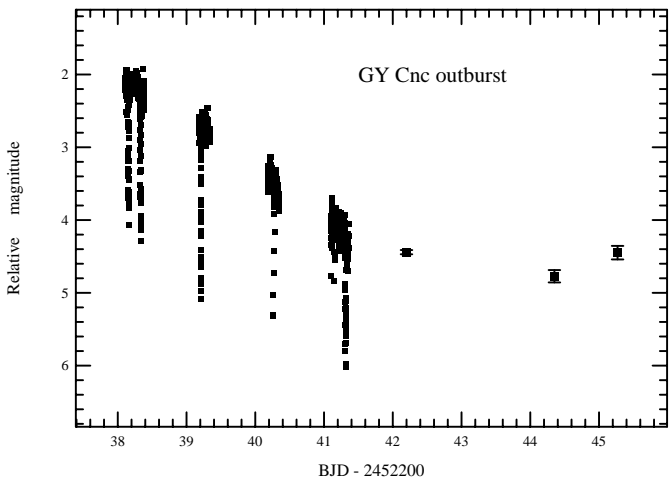}
  \end{center}
  \caption{Overall light curve of the 2001 November outburst.  The last
  three points and error bars represent nightly averaged magnitudes and
  errors of snapshot observations.}
  \label{fig:burst}
\end{figure}

   The 2001 November outburst was detected by J. Gunther and E. Morillon
(AFOEV), and was reported to VSNET.\footnote{
$\langle$http://www.kusastro.kyoto-u.ac.jp/vsnet/$\rangle$}
Following this alert, we conducted time-series CCD photometry on five
consecutive nights (November 24 through 28), which entirely covered
the declining branch of the outburst.  We obtained additional snapshot
photometry on November 30 and December 1, thereby confirming that the
object almost reached quiescence.  The observations were done using
an unfiltered ST-7E camera (system close to R$_{\rm c}$) attached to
the Meade 25-cm Schmidt-Cassegrain telescope.
The exposure time was 30 s.  The images were dark-subtracted,
flat-fielded, and analyzed using the Java$^{\rm TM}$-based PSF
photometry package developed by on of the authors (TK).  The differential
magnitudes of the variable were measured against GSC 1404.1852 (Tycho-2
$V$-magnitude 11.08), whose constancy was confirmed by comparison with
GSC 1404.778 (Tycho-2 $V$-magnitude 11.12).  Barycentric corrections to
the observed times were applied before the following analysis.  The log
of observations is summarized in table \ref{tab:log}.  The overall
light curve is shown in figure \ref{fig:burst}.

\begin{table*}
\caption{Log of observations.}\label{tab:log}
\begin{center}
\begin{tabular}{lcccc}
\hline\hline
Date      & BJD$^*$ (start--end) & N$^\dagger$ & Mag$^\ddagger$ &
            Error$^\S$ \\
\hline
2001 November 24 & 52238.110--52238.381 & 598 & 2.243 & 0.011 \\
2001 November 25 & 52239.179--52239.344 & 298 & 2.768 & 0.015 \\
2001 November 26 & 52240.179--52240.350 & 302 & 3.487 & 0.011 \\
2001 November 27 & 52241.103--52241.357 & 525 & 4.169 & 0.010 \\
2001 November 28 & 52242.154--52242.381 & 125 & 4.441 & 0.029 \\
2001 November 30 & 52244.351--52244.356 &  11 & 4.772 & 0.085 \\
2001 December  1 & 52245.129--52245.375 &  88 & 4.448 & 0.093 \\
\hline
 \multicolumn{5}{l}{$^*$ BJD$-$2400000.} \\
 \multicolumn{5}{l}{$^\dagger$ Number of frames.} \\
 \multicolumn{5}{l}{$^\ddagger$ Averaged magnitude relative to GSC 1404.1852.} \\
 \multicolumn{5}{l}{$^\S$ Standard error of the averaged magnitude.} \\
\end{tabular}
\end{center}
\end{table*}

\section{Results and Discussion}

\subsection{Outburst Frequency}\label{sec:freq}

\begin{figure}
  \begin{center}
    \FigureFile(88mm,60mm){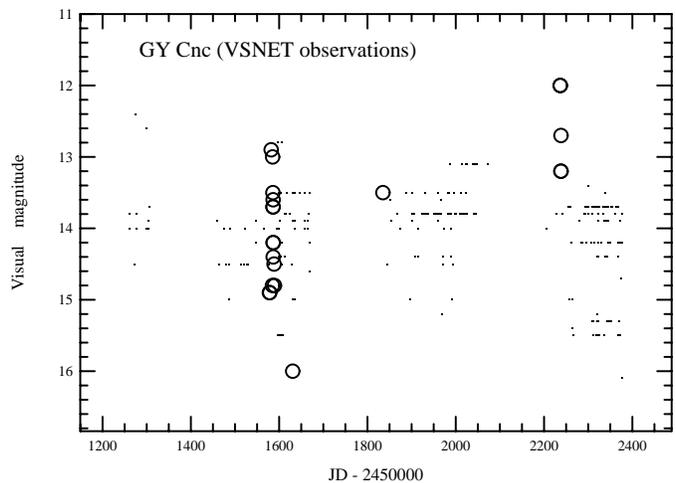}
  \end{center}
  \caption{Long-term light curve from observations reported to VSNET.
  Large open circles and small dots represent positive observations
  and upper limits, respectively.
  }
  \label{fig:vis}
\end{figure}

   There are three known outbursts (2002 February, 2000 October, 2001
November) between 1999 March and 2002 April, when the object was
regularly monitored by VSNET observers.  The light curve drawn from
the available observations is shown in figure \ref{fig:vis}.
Although there are unavoidable gaps in observations (GY Cnc is located
close to the ecliptic), the lack of outbursts immediately following the
2000 October and 2001 November outbursts strongly suggests that the
typical interval of outbursts is an order of 200--300 d.  This value
is one of the longest recurrence times among dwarf novae with similar
orbital periods (\cite{sim00chuma}; \cite{can88outburst}).

\subsection{Eclipse Ephemeris}\label{sec:ephem}

   After subtracting linear declining trends, the mid-eclipse times were
determined by minimizing the dispersions of eclipse light
curves folded at the mid-eclipse times.  The error of eclipse times were
estimated using the Lafler--Kinman class of methods, as applied by
\citet{fer89error}.  The validity of the estimated errors has been
confirmed by an application to different ranges (in eclipse depth) of
the data.\footnote{As discussed in subsection \ref{sec:eclprof},
  the eclipse profiles were slightly asymmetric, possibly causing a slight
  systematic error.
  The error estimates should therefore be treated as a statistical measure
  of the observational errors.
}
Table \ref{tab:eclmin} summarizes the observed times of the
eclipses (labeled as ``this work"), together with the published eclipse
times summarized in \citet{kat00gycnc}.  The times from \citet{kat00gycnc}
were converted into the Barycentric Julian Date (BJD) system,
common to the present observation.
The cycle count ($E$) follows the definition by \citet{kat00gycnc}.

\begin{table}
\caption{Eclipses and $O-C$'s of GY Cnc.}\label{tab:eclmin}
\begin{center}
\begin{tabular}{lrrrc}
\hline\hline
Eclipse$^*$ & Error$^\dagger$ & $E$$^\ddagger$ & $O-C$$^\S$
                 & Ref.$^\|$ \\
\hline
51581.8263  & $\cdots$ & $-$25 & $-$34 & 1 \\
51582.0017  & $\cdots$ & $-$24 & $-$39 & 1 \\
51585.6861  & $\cdots$ &   -3  & $-$28 & 1 \\
51586.21255 & $\cdots$ &    0  & $-$16 & 2 \\
51586.38834 & $\cdots$ &    1  &    19 & 3 \\
51586.56372 & $\cdots$ &    2  &    13 & 3 \\
51586.91465 & $\cdots$ &    4  &    17 & 2 \\
51587.96813 & $\cdots$ &   10  &   100 & 2 \\
51588.31802 & $\cdots$ &   12  &     0 & 2 \\
51589.8969  & $\cdots$ &   21  & $-$10 & 1 \\
51590.07239 & $\cdots$ &   22  &  $-$5 & 2 \\
51590.7742  & $\cdots$ &   26  &  $-$1 & 1 \\
51590.9496  & $\cdots$ &   27  &  $-$5 & 1 \\
51590.94997 & $\cdots$ &   27  &     3 & 2 \\
51591.12457 & $\cdots$ &   28  &  $-$5 & 2 \\
51599.7219  & $\cdots$ &   77  &    12 & 1 \\
52238.15700 &     4    & 3716  &  $-$6 & 4 \\
52238.33258 &     3    & 3717  &     7 & 4 \\
52239.20975 &     5    & 3722  &     4 & 4 \\
52241.31496 &     9    & 3734  &  $-$1 & 4 \\
\hline
 \multicolumn{5}{l}{$^*$ Eclipse center.  BJD$-$2400000.} \\
 \multicolumn{5}{l}{$^\dagger$ Estimated error in 10$^{-5}$ d.} \\
 \multicolumn{5}{l}{$^\ddagger$ Cycle count.} \\
 \multicolumn{5}{l}{$^\S$ Against equation (\ref{equ:reg}).
                           Unit in 10$^{-5}$ d.} \\
 \multicolumn{5}{l}{$^\|$ Reference.} \\
 \multicolumn{5}{l}{\phantom{$^\|$} 1: \citet{gan00gycnc},
                    2: \citet{kat00gycnc}}, \\
 \multicolumn{5}{l}{\phantom{$^\|$}  3: T. Vanmunster (vsnet-alert 4210),
                    4: this work} \\
\end{tabular}
\end{center}
\end{table}

   Using these eclipse times, we obtained the following linear ephemeris.
The orbital phases used in the following figures and discussions are based
on the following equation:

\begin{equation}
\rm{BJD_{min}} = 2451586.21271(8) + 0.17544251(5) $E$. \label{equ:reg}
\end{equation}

\subsection{Outburst Light Curve}\label{sec:lc}

\begin{figure}
  \begin{center}
    \FigureFile(88mm,60mm){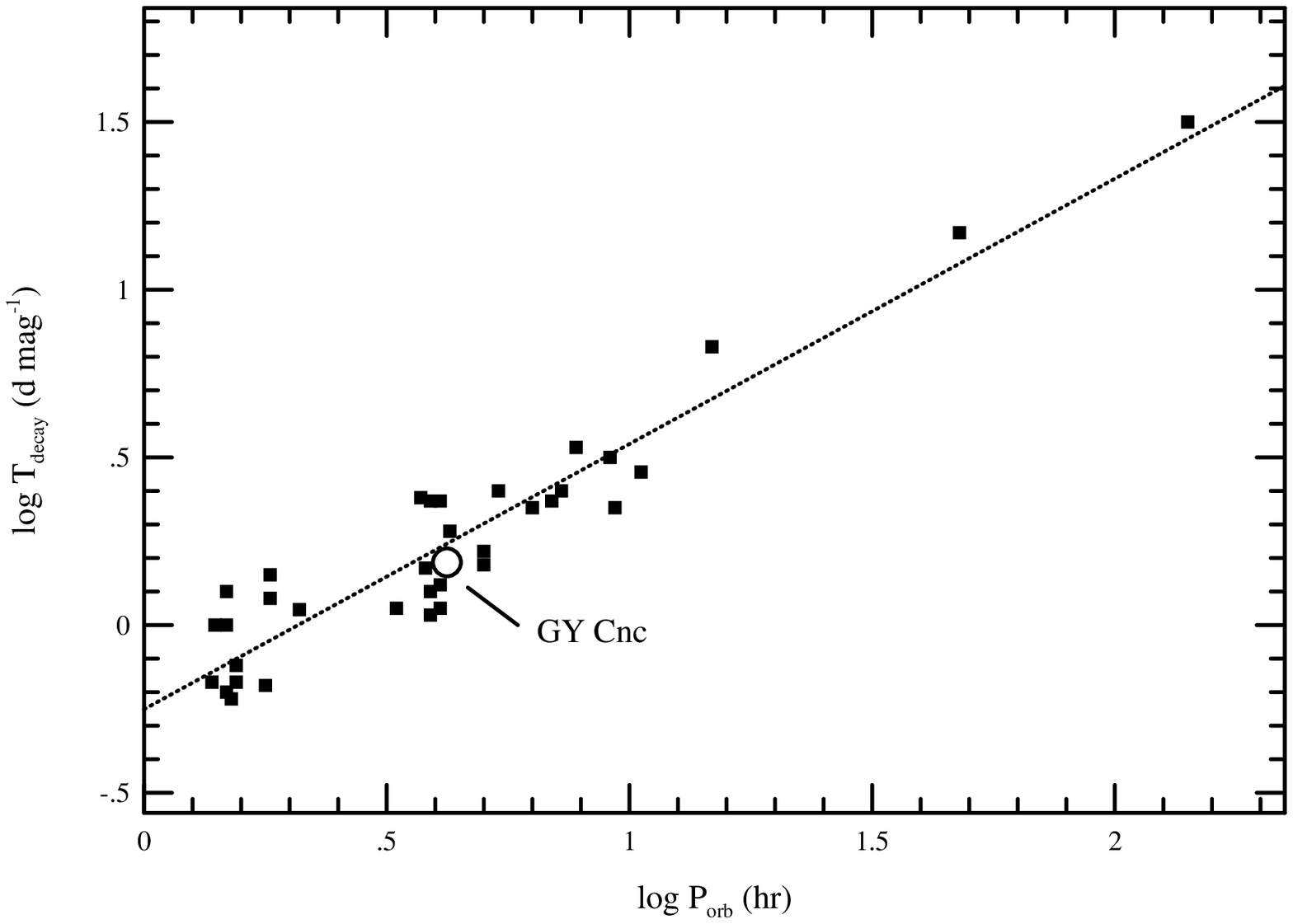}
  \end{center}
  \caption{Relation between orbital periods and rates of decline of
  well-observed dwarf novae.  The data (dots) are mainly taken from
  a compilation by \citet{pat97crboo} based on \authorcite{war87CVabsmag}
  (\yearcite{war87CVabsmag},\yearcite{war95book}),
  excluding an helium system CR Boo, and including newly measured systems
  (DX And: \cite{kat01dxand}; WX Cet \cite{kat01wxcet};
  IR Com: \cite{kat02ircom}).  The dotted line indicates a linear regression
  ($\log T_{\rm decay} = -0.25 + 0.79 \log P_{\rm orb}$) from the
  available data.  The location of GY Cnc is shown with an open circle.
  }
  \label{fig:bailey}
\end{figure}

   As shown in figure \ref{fig:burst}, the object showed an almost linear
decline during the first four nights.  The mean rate of decline was
0.65 mag d$^{-1}$.  This value almost exactly fits the well-known
Bailey's relation (e.g. \cite{BaileyRelation})
between orbital periods and rates of decline of dwarf novae
(figure \ref{fig:bailey}).
This finding indicates that the fading part of the GY Cnc is
indistinguishable from those of typical dwarf novae.  The rising
portion of the outburst was not observed; the last negative observation
reported to VSNET was made 8 d before the detection of the outburst.

\subsection{Orbital Modulations}

\begin{figure}
  \begin{center}
    \FigureFile(88mm,60mm){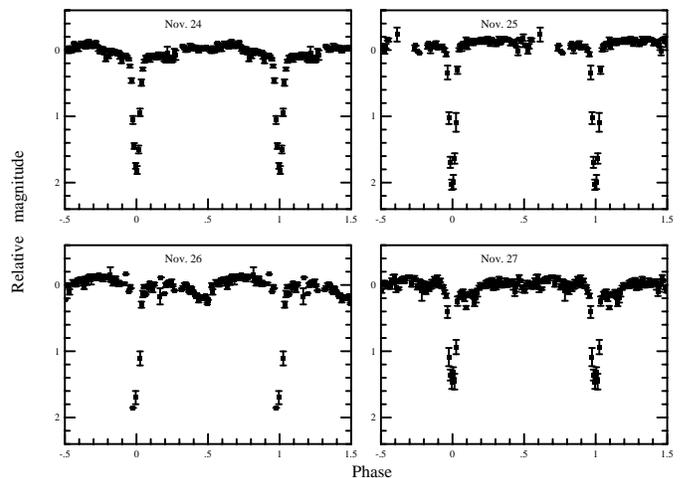}
  \end{center}
  \caption{Phase-averaged light curves of GY Cnc during the decline phase
  (2001 November 24--27).  Each point represents an average of 0.01 phase
  bin.}
  \label{fig:orbital}
\end{figure}

\begin{figure}
  \begin{center}
    \FigureFile(88mm,60mm){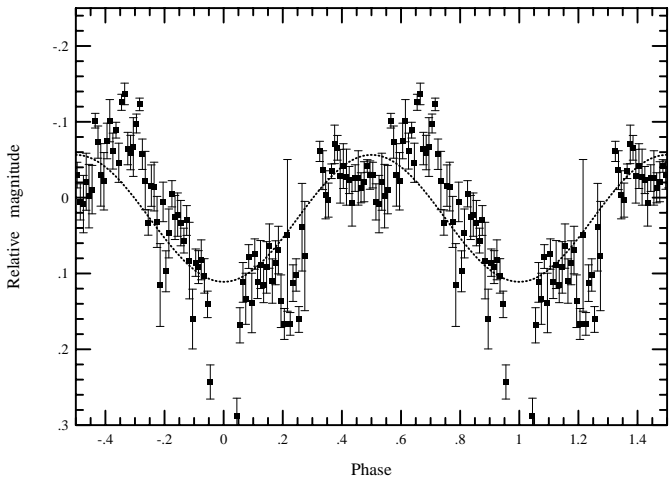}
  \end{center}
  \caption{Reflection effect and possible phase 0.5 shallow dip on November 24.
  The dashed lines represent the best sinusoidal fit to the light curve
  outside the eclipses: $\Delta mag = 0.027 + 0.084 \cos(phase)$.  Phases
  between $-$0.08 and 0.08 correspond to an eclipse.
  }
  \label{fig:reflect}
\end{figure}

   Figure \ref{fig:orbital} shows phase-averaged light curves during the
decline phase (2001 November 24--27), after subtracting the linear decline
(0.65 mag d$^{-1}$).  Orbital phases were calculated against equation
(\ref{equ:reg}).  On November 24, the light curve showed a moderate
($\sim$0.1--0.2 mag) reflection effect and a possible shallow secondary
eclipse around phase 0.5 (Figure \ref{fig:reflect}).
These findings indicate a substantial amount
of irradiation from the outbursting disk on the secondary star.  The
presence of the possible phase 0.5 eclipse suggests that the heated
surface of the secondary was partially eclipsed by the optically thick
accretion disk, as in IP Peg \citep{lit01ippegmirroreclipse}.
GY Cnc in outburst thus would be an excellent target for detecting
``mirror eclipses" of profiles of emission lines.

On the second night (November 25), the light
curve outside the eclipses became flatter, indicating a lesser degree
of irradiation as the outburst faded.  On the third night (November 26),
a hump (orbital phase 0.6--0.9) became detectable, but it is far less
prominent than in other high-inclination dwarf novae
(e.g. U Gem \cite{krz65ugem}).
There may be a dip-like structure around phase 0.4--0.5, whose origin
is unknown.  On the fourth night (November 27), a ``shoulder" became
evident between orbital phases 0--0.2, which is likely to be associated
with the hot spot.

\subsection{Eclipse Profile}\label{sec:eclprof}

\begin{figure*}
  \begin{center}
    \FigureFile(130mm,90mm){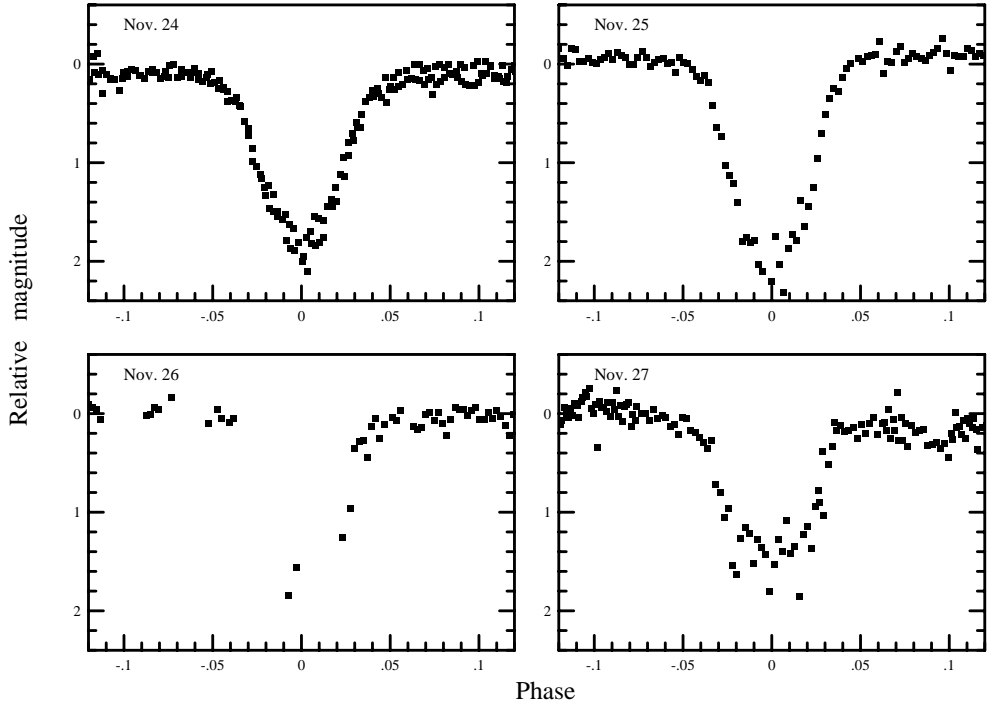}
  \end{center}
  \caption{Eclipse profiles.  Each point represent individual observations.
  }
  \label{fig:ecl}
\end{figure*}

\begin{figure}
  \begin{center}
    \FigureFile(88mm,120mm){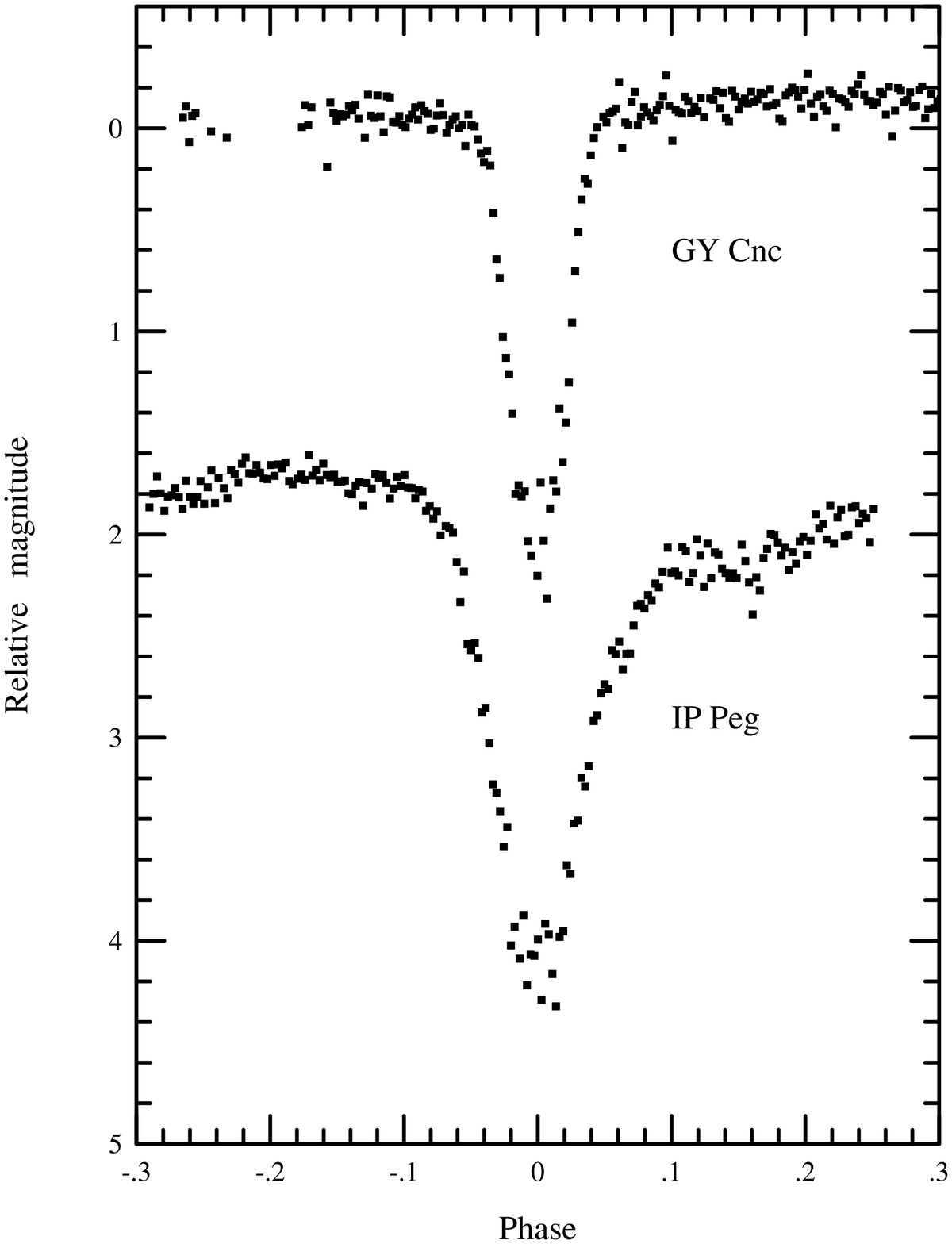}
  \end{center}
  \caption{Comparison of eclipses between GY Cnc and IP Peg, observed
  $\sim$2 mag above the quiescent level.  The observation of IP Peg
  was done on 1995 December 23 (during the fading branch of an outburst).
  A pre-eclipse orbital hump and a slower egress are already prominent
  in IP Peg, while the light curve outside the eclipse is virtually
  flat in GY Cnc.
  }
  \label{fig:comp}
\end{figure}

   Figure \ref{fig:ecl} shows enlarged profiles of the eclipses.  On the
third night (November 26), only the egress part of an eclipse was observed.
On other nights, entire phases of eclipses were observed.  On the first
two nights (November 24 and 25), the eclipses were ``V"-shaped, as is
usual for a high-inclination dwarf nova in outburst (cf. \cite{bap00exdra}).
However, GY Cnc shows a some degree of peculiarity in that at least some
of eclipses show almost identical ingress and egress [For example, the mean
rates of variation between $0.01 < |phase| < 0.04$ are 50 and 52 mag d$^{-1}$
for ingress and egress on November 24, respectively.  On November 25,
the corresponding mean rates are 68 and 64 mag d$^{-1}$.  Estimated errors
are 2 mag d$^{-1}$].  This is contrary to what
is expected and predicted for an effect of the presence of a hot spot
\citep{sha00gycnc}.  This finding suggests that the accretion disk
in outbursting GY Cnc has a brighter portion in the trailing direction
(i.e. the opposite direction to the expected hot spot).  Figure
\ref{fig:comp} shows a comparison eclipses between GY Cnc and IP Peg,
observed $\sim$2 mag above the quiescent level.  A pre-eclipse orbital
hump and a slower egress are already prominent in IP Peg, while the light
curve outside the eclipse is virtually flat in GY Cnc.

   Table \ref{tab:width} lists eclipse widths, which are defined as the
durations when the object is more than 0.2 mag fainter than the mean
magnitudes outside the eclipses.  The data on November 26 were not used
because they lack ingress/egress observations.  The width of eclipse
dramatically decreased between November 24 and 25, which indicates a
dramatic shrinkage of the luminous part of the accretion disk.  The decrease
of the width stopped between November 25 and 27, which suggests that
the inner luminous part of the accretion disk disappeared, i.e. almost
the entire disk again became in low temperature state
(cf. \cite{osa96review}).  This finding is compatible with the observation
that the object was only $\sim$0.5 mag above  its quiescent magnitude
on November 27 (cf. Figure \ref{fig:burst}).

\begin{table}
\caption{Eclipse width.}\label{tab:width}
\begin{center}
\begin{tabular}{lccc}
\hline\hline
Date             & $\phi_{\rm in}$$^*$ & $\phi_{\rm eg}$$^*$
                 & Eclipse width \\
\hline
2001 November 24 & $-$0.049 & 0.052 & 0.101 \\
2001 November 25 & $-$0.039 & 0.038 & 0.077 \\
2001 November 27 & $-$0.042 & 0.042 & 0.084 \\
\hline
 \multicolumn{4}{l}{$^*$ Phases of eclipse ingress and egress, which are} \\
 \multicolumn{4}{l}{\phantom{$^*$} defined as the point when the object is} \\
 \multicolumn{4}{l}{\phantom{$^*$} 0.2 mag below the mean magnitude outside the} \\
 \multicolumn{4}{l}{\phantom{$^*$} eclipses.} \\
\end{tabular}
\end{center}
\end{table}

\subsection{Nature of GY Cnc}

   As shown in above subsections, orbital humps in GY Cnc were less
prominent than in other high-inclination dwarf novae even during the
later stage of an outburst.  In conjunction with the absence of asymmetric
features of eclipses which are expected from a hot spot (subsection
\ref{sec:eclprof}), we conclude that the hot spot is weaker in GY Cnc
than in other dwarf novae.  This is in agreement with the finding
by \citet{sha00gycnc}, who studied the object in post-outburst
quiescence.  Such a feature may be a consequence of a smaller mass-transfer
rate ($\dot{M}$) than in other dwarf novae above the period gap.
Although the low occurrence of outbursts (subsection \ref{sec:freq}),
may be explained by the assumption of a small mass-transfer rate
or a low quiescent viscosity \citep{ich94cycle},
low-$\dot{M}$ systems above the period gap tend to show slowly rising
long outbursts (cf. CH UMa, \citet{sim00chuma}; DX And \citet{sim00dxand}).
This is a natural consequence of disk-instabilities arising from the
inner accretion disk (cf. \cite{sma84DI}; see also discussions in
\citet{sim00chuma}).  In contrast to DX And and CH UMa, no known outbursts
of GY Cnc are slowly rising, long outbursts.

In conjunction with a relatively high $L_{\rm X}/L_{\rm opt}$
\citep{bad98RASSID}, we suggest an alternative explanation.
As proposed in \citet{kat02ircom}, two dwarf novae (HT Cas and IR Com)
show common properties (whose origin is not yet well understood):
relatively high $L_{\rm X}/L_{\rm opt}$,
low frequency of outbursts, and little evidence of orbital humps, all
of which are common to GY Cnc.  We suspect that GY Cnc may be the
first counterpart of HT Cas or IR Com above the period gap.

Although there is no strongly supporting observational evidence,
a truncation of the inner accretion disk by the magnetic field of
the white dwarf, as in intermediate polars (IPs), could also be a viable
explantion.  In such a condition, the truncation effectively suppresses
disk instabilities in the inner accretion disk, thereby lengthening
outburst intervals \citep{ang89DNoutburstmagnetic}.
Some IPs show dwarf nova-type outbursts, the best known examples being
DO Dra\footnote{DO Dra has been the source of some confusion
  regarding its designation. Some authors call the same variable YY Dra,
  however, we use the designation DO Dra, following the official nomenclature
  by the GCVS (\cite{NameList67}; \cite{kho88dodra}.}
(\cite{wen83dodra}; \cite{szk02dodra}), TV Col (\cite{szk84tvcolflare};
\cite{hel93tvcolperiods}), EX Hya \citep{hel89exhya} and
HT Cam (\cite{ish02htcam}; \cite{kem02htcam}).\footnote{
  Other IPs showing outbursts include GK Per, XY Ari and V1223 Sgr.}
Most of them show brief (and often small) outbursts, which often show
a precipitous decline during its final stage [e.g. HT Cam showed a decline
rate of 4 mag d$^{-1}$ during its late stage of an outburst
\citep{ish02htcam}].  Such presence of a precipitous decline is naturally
explained if the inner part of the accretion disk is magnetically
truncated \citep{ish02htcam}.  From this viewpoint, the lack of departure
in GY Cnc from the Bailey's relation throughout the outburst decline
(subsection \ref{sec:lc}) seems to less favor the possibility of the
IP interpretation.
This IP possibility could be more directly tested by future attempts to
directly detect spin pulses or X-ray modulations arising from a magnetized
white dwarf.

\section{Conclusion}

   We observed the ROSAT-selected eclipsing dwarf nova GY Cnc
(=RX J0909.8+1849) during the 2001 November outburst.  We refined
the orbital period to be 0.17544251(5) d.  The fading portion of
the outburst was indistinguishable from those of typical dwarf novae
with similar orbital periods.  However, the orbital light curves
and eclipse profiles show a lesser signature of orbital humps
(or a hot spot) than in usual dwarf novae with similar orbital periods.
We suspect that GY Cnc may be the first above-the-gap counterpart of
unusual eclipsing dwarf novae HT Cas and IR Com.

\vskip 3mm

We are grateful to many VSNET observers who have reported vital observations.
This work is partly supported by a grant-in aid (13640239) from the
Japanese Ministry of Education, Culture, Sports, Science and Technology.
Part of this work is supported by a Research Fellowship of the
Japan Society for the Promotion of Science for Young Scientists (MU).

\end{document}